# Tests of Cold Atom Clock in Orbit

Liang Liu[1*], Desheng Lü[1*], Weibiao Chen[2*], Tang Li[1], Qiuzhi Qu[1], Bin Wang[1], Lin Li[1], Wei Ren[1], Zuoren Dong[2], Jianbo Zhao[1], Wenbing Xia[2], Xin Zhao[1], Jingwei Ji[1], Meifeng Ye[1], Yanguang Sun[2], Yuanyuan Yao[1], Dan Song[1], Zhaogang Liang[1], Shanjiang Hu[2], Dunhe Yu[2], Xia Hou[2], Wei Shi[2], Huaguo Zang[2], Jingfeng Xiang[1], Xiangkai Peng[1], Yuzhu Wang[1]

[1]Key Laboratory of Quantum Optics, Shanghai Institute of Optics and Fine Mechanics, Chinese Academy of Sciences, Shanghai 201800, China.

[2]Research Center of Space Laser Information Technology, Shanghai Institute of Optics and Fine Mechanics, Chinese Academy of Sciences, Shanghai 201800, China.

*Correspondence to: liang.liu@siom.ac.cn, dslv@siom.ac.cn, wbchen@mail.shcnc.ac.cn

**Abstract**: Since the atomic clock was invented, its performance has been improved for one digit every decade until 90s of last century when the traditional atomic clock almost reached its limit. With laser cooled atoms, the performance can be further improved, and nowadays the cold atom based clocks are widely used as primary frequency standards. Such a kind of cold atom clocks has great applications in space. This paper presents the design and tests of a cold atom clock (CAC) operating in space. In microgravity, the atoms are cooled, trapped, launched and finally detected after being interrogated by microwave field with Ramsey method. The results of laser cooling of atoms in microgravity in orbit are presented and compared with that on ground for the first time. That the full width at half maximum (FWHM) of obtained central Ramsey fringes varies linearly with launching velocity of cold atoms shows the effects of microgravity. With appropriate parameters, a closed-loop locking of the CAC is realized in orbit and the estimated short term frequency stability of $3.0 \times 10^{-13}/\sqrt{\tau}$ has been reached.

**One Sentence Summary:** The test results of a space cold atom clock in orbit are presented.

The atomic clock has great applications in time service, satellite navigation, tests of fundamental physics, and also in many industries. Traditional atomic clocks include rubidium clock, hydrogen maser and cesium beam clock. Different kind of clocks serves different applications, but the performance of those traditional clocks has almost reached their limits by now.

Laser cooling of atoms provides a new approach to improve the performance of atomic clocks further (*1*). The atoms are first cooled by lasers, and then interrogated by microwave field typically with Ramsey method. The width of the central Ramsey fringe for a cold atom clock is almost 2 order narrower than that for the hot atom counterpart. The most successful cold atom clock (CAC) is so-called atomic fountain, whose uncertainty can reach around 2E-16 (*2-4*). The fountain clock is now the most accurate atomic clock and widely used as the primary frequency standard. In a typical fountain clock, atoms are first cooled by lasers, and then launched upwards by moving molasses. Due to the gravity, the launched cold atoms drop after they reach the apogee. The upward and downward movement of cold atoms leads to twice interaction with the microwave field in a cavity, and thus give the Ramsey signal.

In space, the microgravity offers a new opportunity for the CAC. In microgravity, the launched cold atoms move in a uniform velocity, and thus the interrogation time is determined by the launching velocity, and thus the Ramsey width can be further reduced at a reasonable device size.

The space clock has many applications in fundamental physics. For example, the atomic clock in orbit can be used to measure the red shift of frequency (*5*). Large scale physics requires a precision measurement of time and distance by space clocks. The high performance space atomic clocks can also help to search dark matter (*6*), to detect gravitational wave (*7*), and etc. and play an important role in deep space navigation. Besides, the cold atom physics in space is the base of many other techniques required for cold atom interferometry, optical clocks, and cold atom sensors.

The experiments related to cold atoms in microgravity have been successfully performed in drop tower, parabolic flight or sounding rocket (*8-10*). However, with those microgravity methods, it is impossible to perform a long-term continuous operation, which is a requirement for an atomic clock. Only flying in orbit can offer a long-term microgravity environment so that the space CAC can operate continuously. Moreover, such in-orbit operation can accumulate orbital environment data specified for cold atom physics in space.

To operate such a CAC in orbit has great challenges. First, due to the limited space resources, the weight, volume and power consumption must be greatly reduced compared to the ground fountain clock. Second, the space CAC must pass all strict mechanical, thermal and electromagnetic compatibility test. Third, all operations of the CAC must be automatic and all units must be maintained without any hand-adjustment. Fourth, the CAC must tolerate the orbital environment, such as the variation of earth magnetic field and high energy particles. And finally, the CAC must be designed to work in microgravity.

Several projects on space CACs have been proposed in the last decade. For example, the ACES mission aims mainly to operate a cold atom clock PHARAO in the International Space Station, and to measure the gravitational red shift of the clock frequency, test Lorentz invariance and search the variations of fundamental physical constants (*11-16*). We started a mission called Cold Atom Clock Experiment in Space (CACES) in 2011 under the support of China Manned Space Program. The aims of the CACES are
1. to test laser cooling of atoms in orbit,
2. to test the launching and motion of cold atoms at a very low velocity in microgravity,
3. to test the interrogation of cold atoms in Ramsey microwave fields in microgravity,
4. to test the key technologies in orbit, including diode lasers and their frequency stabilization, optical system, fiber couplers, vacuum, temperature control in microgravity, and etc.,
5. to estimate the performance of a CAC in orbit,
6. to accumulate data of the orbital environment for cold atom physics.

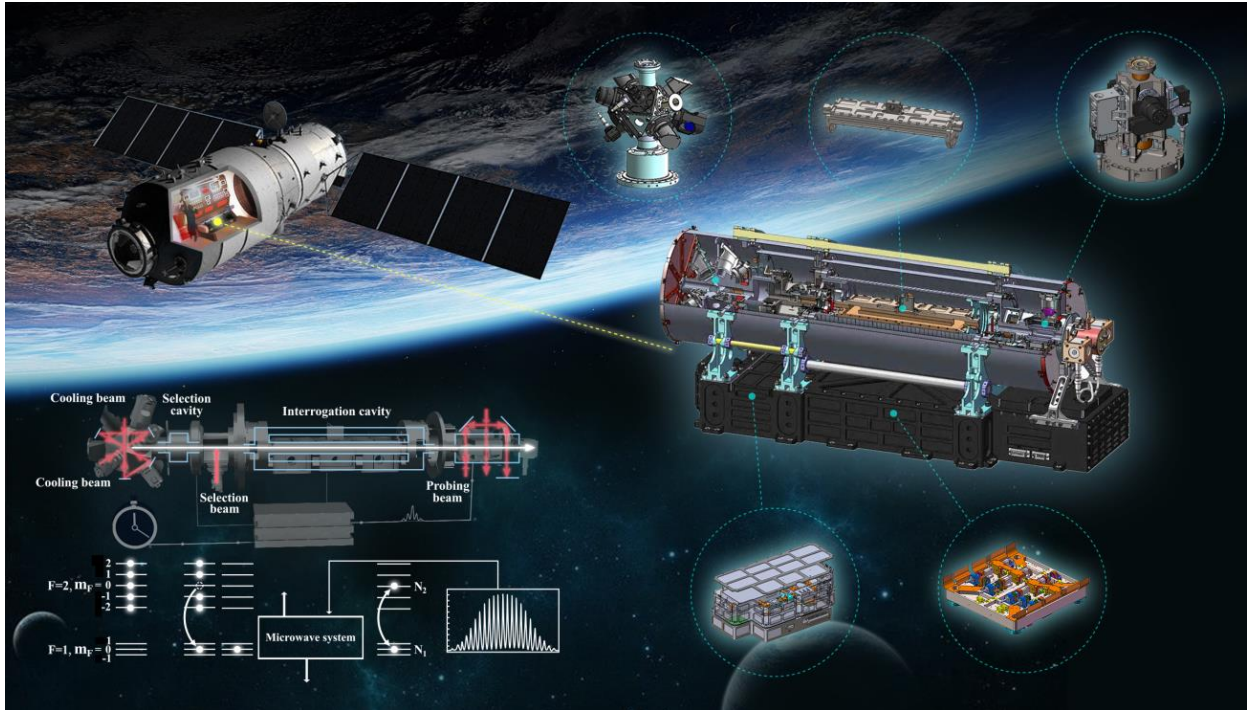

Fig. 1. Principle and structure of the space CAC located inside the spacecraft Tiangong-2. The laser cooling zone is a MOT with folded beam design. The microwave cavity is a ring cavity used for the microwave signal to interrogate cold atoms. In the detection zone, the cold atoms on both hyperfine states are detected. The optical unit includes all optics and lasers offering the requirements of cooling, state selection and detection of cold atoms. The microwave unit is a synthesizer which converts 5MHz to 6.8GHz. The functions of the control unit not only determine the CAC timing sequence, but also are used to communicate with the control center of the spacecraft.

The principle of a CAC operating in orbit is shown in Fig. 1. Rubidium is used in our CAC mainly due to its relatively higher melting temperature, compared to cesium, which reduces the design complexity of the atomic source for space application. The atoms are cooled and trapped in a magneto-optical trap (MOT), and then the cooled atoms are launched by moving molasses. During the launching, the cold atoms are further cooled by adiabatic cooling. Because of the microgravity, the cold atoms move in a straight line at a uniform velocity. After state selection, the cold atoms are interrogated by the microwave field, and then detected by the laser excited atomic fluorescence.

A CAC specified for the space applications was designed which consists of four units: physical package, optical bench, microwave source and control electronics, as shown in Fig. 1. The main part of the physical package is a titanium alloy vacuum tube whose vacuum is maintained to be better than $1 \times 10^{-7}$Pa during the test *(17)*. A ring cavity is used for the Ramsey interrogation of cold atoms by the microwave field *(18)*. Three layers of Mumetal are used to shield the magnetic field at the interrogation cavity, but only one layer at the cooling zone. The magnetic field in the interrogation cavity is automatically compensated by regulating the current in the C-field coil through a servo loop to keep total magnetic field stable during the motion of the spacecraft in orbit *(19)*.

Together with a pair of anti-Helmholtz coils, two laser beams are folded to form a compact MOT in order to reduce the requirement of laser power *(20)*. The frequency of the two beams, one is folded such that all beams intersected at the cooling region in the MOT are transmitted forward and another backward, can be controlled separately to form a moving molasses. The cold atoms can be launched by the moving molasses at a velocity

$$v = (\omega_1 - \omega_2)/k , \tag{1}$$

where $\omega_1$, $\omega_2$ are the frequency of the two laser beams respectively and $k$ is the wave vector.

The state selection is fulfilled by a combination of microwave excitation and laser pushing method. After being cooled, the atoms are concentrated at the state $|F = 2>$, evenly distributed at 5 magnetic sub-states. The microwave power in the state selection cavity is adjusted such that the atoms at $|F = 2, m_F = 0>$ are pumped to $|F = 1, m_F = 0>$ at an efficiency of almost 100%, and then a laser beam pushes all other atoms at $|F = 2, m_F \neq 0>$ states away, and the remained atoms are used for the microwave interrogation.

The Ramsey interrogation of a space CAC requires two cavities separated by a distance $D$, leading to the interrogation time $T=D/v$ for cold atoms with velocity $v$ in microgravity, and the width of central Ramsey fringe is directly related to the velocity as (*21*)

$$\Delta \propto v/2D . \tag{2}$$

Considering the dead time of the clock cycle combined with the size limitation, $D = 217$ mm in our space CAC.

The optical bench is the key unit for the space CAC. The bench includes 3 diode lasers, 5 AOMs, 5 fiber couplers, 2 sets of rubidium cells and detectors for saturation spectroscopy and optical components such as reflectors, beam splitters, half and quarter wave plates, and optical isolators. In order to keep the output power of each laser beam unchanged for a long period, the bench is carefully designed to keep temperature and mechanics stable, and all optical units are glue-bound to a SiC-reinforced AI–Si alloy plate (*22*). With such a design, the bench has been used to work on the engineering model of the space CAC for more than 4 years, and its performance is still kept same as before without any adjustment. The flight model has been working for almost one year in orbit after launched, and the performance is still kept as designed. In the bench, the saturation spectroscopy is used to lock the frequency of lasers to the transitions of the rubidium D2 line. The lasers can be kept locking for months unless they are intentionally interrupted. Besides, once the lasers are unlocked, they can be re-locked automatically in less than a few minutes.

The microwave source is based on the frequency synthesis of an ultra-stable 5 MHz quartz oscillator (OCXO) up to 6.834 GHz. By using a two-step phase-lock-loop (PLL) architecture, the output microwave signal benefits from both ultra-high close-in phase stability of the reference quartz oscillator and low phase noise level of the PLL at frequencies far from the carrier. At frequencies near from the carrier, the phase noise level of the microwave signal is mainly determined by the 5 MHz oscillator while the additive phase noise of frequency synthesizer is 10 dB lower. This is important for the space CAC with long clock period to suppress the influence of Dick effect.

The control electronic unit consists of two parts: the microcontroller unit and the FPGA unit. The microcontroller unit manages the data communication and calculation. The timing sequence for the space CAC in orbit is controlled by the FPGA unit, and all parameters in the sequence can be adjusted remotely.

We have tested the CACES setup in the laboratory and all performance satisfies the designed requirements (23). The setup passed all mechanical, thermal and electromagnetic compatibility test required by China Manned Space Program. It was launched on September 15, 2016 with Tiangong-2 into orbit and started the test the next day. Since then, the CACES has been working in orbit under management of the Tiangong-2. The almost infinite microgravity time in orbit allows us to test the space CAC relatively easily.

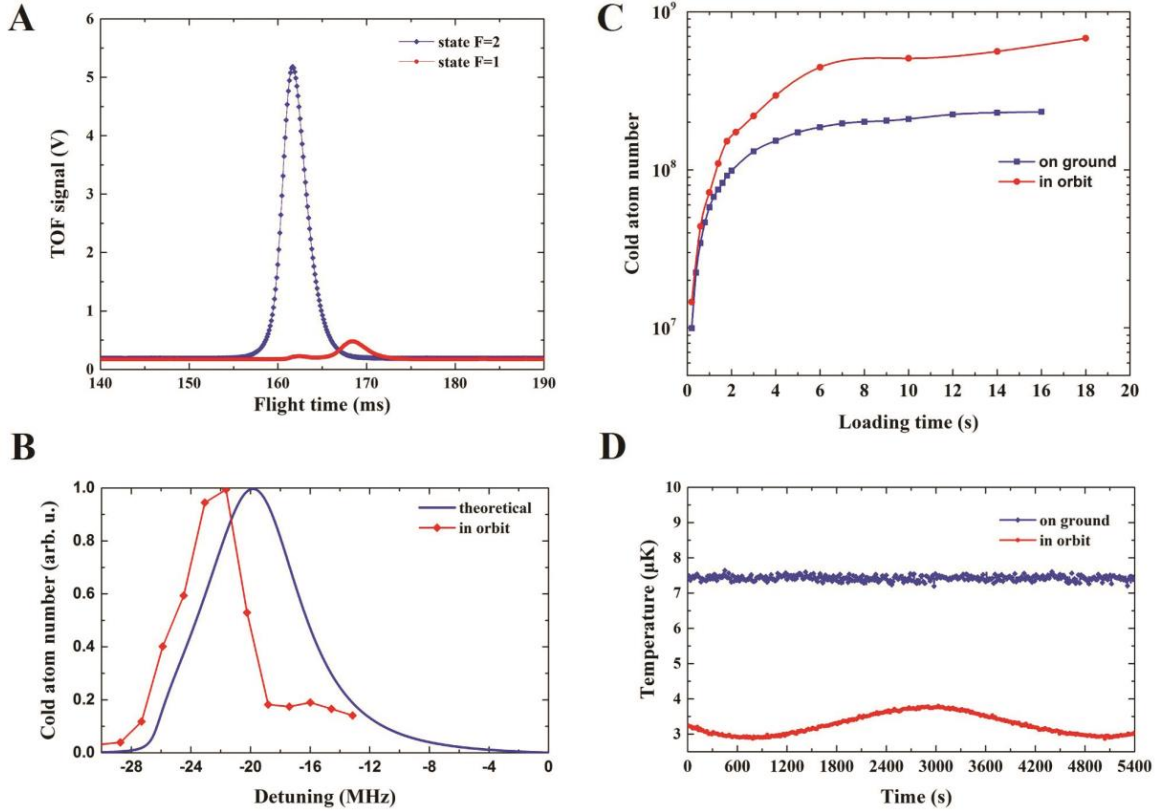

Fig. 2. Laser cooling of atoms in orbit. (**A**) Typical TOF signal. The red curve corresponds to the fluorescence from state $|F = 1, m_F = 0>$, and the blue from state $|F = 2, m_F = 0>$. Launching velocity is 4 m/s. (**B**) Relative number of cold atoms in the MOT vs the detuning of cooling lasers. (**C**) Loading of cold atoms in the MOT. The red curve corresponds to the loading in orbit, and blue one on ground at same conditions. Here the detuning and intensity of each cooling laser beam are 22.0 MHz and 3.8 mW/cm², respectively. (**D**) Temperature of cold atoms in orbit vs the measuring time. The temperature is measured after post-cooling and the parameters are optimized to maximize the number of cold atoms and minimize the temperature.

Fig. 2 presents the typical results of laser cooling of atoms in orbit. Fig. 2(**A**) gives time of flight (TOF) signal at detection region. The cold atoms are launched by moving molasses at a velocity given by Eq. (1), and detected by laser exciting the atomic fluorescence. The number of cold atoms in both $|F = 1>$ and $|F = 2>$ of ground states is detected, and the transition probability can be obtained as

$$p = \frac{N_2}{N_1 + N_2} \tag{3}$$

where, $N_1$ and $N_2$ are the number of the cold atoms at the state $|F = 1 >$ and $|F = 2 >$ respectively. With such a detection, the influence of the atomic number fluctuation is greatly reduced.

We have tested the laser cooling and loading of the atoms in a MOT in orbit. Fig. 2(**B**) gives the cold atom number in the MOT vs frequency detuning of the cooling laser over the atomic transition $|F = 2 \to F' = 3 >$. The result shows that at the detuning around -22 MHz the cooling has the highest efficiency, which is in agreement with the theory (*24*). The loading of cold atoms in microgravity was first studied by Kulas and *et. al.*, but no comparison between ground and microgravity was given (*10*). Here we give the first experimental results of the cold atom loading in a MOT in microgravity compared with the results of the ground test as shown in Fig. 2(**C**). The higher loading efficiency of the cold atoms in a MOT in microgravity than on ground is expected mainly due to the less loss of slow atoms during cooling process. Fig. 2(**D**) gives the temperature of cold atoms after post-cooling. The atoms are first cooled and trapped in the MOT, and the post-cooling is realized during the launching of the cold atoms in moving molasses. By adiabatically changing the detuning and power of cooling lasers, a temperature much lower than that of the Doppler cooling can be obtained. The variation of the temperature is mainly due to the change of the near-earth magnetic field resulted from the spacecraft rotating around the earth in orbit. A mean temperature of 3.3 µK with a variation of approximate 0.9µK is obtained. With better magnetic shield, the temperature should be more stable. Fig. 2(**D**) also gives the temperature of cold atoms on ground at similar conditions. The constant earth magnetic field gives stable temperature.

Fig. 3 gives typical signal of microwave interrogation with cold atoms in the Ramsey cavity in orbit. The cavity and microwave frequency are pre-tuned to the resonance with the atomic transition of two ground states $|F = 1, m_F = 0 \to F = 2, m_F = 0 >$. Before entering the interrogation cavity, the cold atoms are prepared at $|F = 1, m_F = 0 >$. By changing the power of microwave injected into the interrogation cavity, the transition probability described in Eq. (3) oscillates, and at certain power, all of the population flips to $|F = 2, m_F = 0 >$. Fig. 3(**A-C**) are the Rabi oscillations of the transition probability between two ground states with different launching velocities in the interrogation cavity and each peak corresponds to $\pi$ oscillation of the two states. Fig. 3 (**D-F**) give typical Ramsey fringes. Unlike in gravity on ground, the width of the central Ramsey fringe in microgravity given by Eq. (2) is linearly related to the launching velocity, as shown in Fig. 4, which shows a clear evidence that the cold atoms move at a uniform velocity in microgravity.

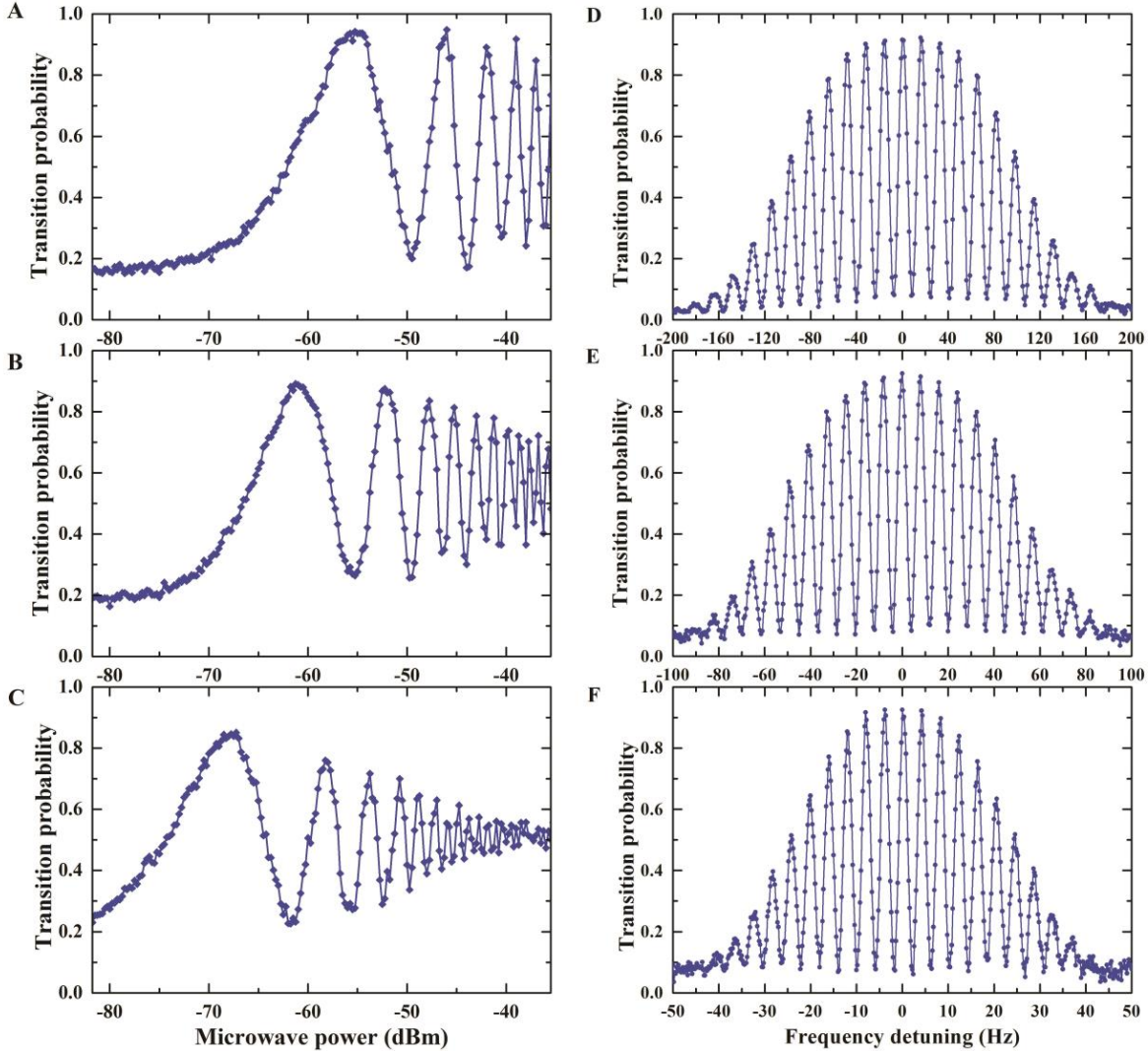

Fig. 3. Microwave interrogation with cold atoms in orbit. The left column gives Rabi oscillations and the right is the Ramsey fringes at different launching velocities of cold atoms. (**A-C**) The typical Rabi oscillations. The microwave frequency is set to resonance with the atomic transition. The launching velocity is (**A**) 4m/s. (**B**) 2m/s and (**C**) 1m/s, respectively. (**D-F**) Typical Ramsey fringes. The microwave power is set to the $\pi$ transition at resonance, and the launching velocity of (**D-F**) is same as (**A-C**), corresponding to the FWHM of the central fringe 7.27 Hz, 3.89 Hz, and 1.80 Hz, respectively.

In order to test the clock performance in orbit, a closed-loop operation has been performed on the CAC by feeding the error signal to the direct digital synthesizer (DDS) of the microwave source. Such operation has been performed continuously for a long period in orbit. Furthermore, a signal-to-noise ratio (SNR) of 440 at the half-maximum point of the central Ramsey fringe with 2 Hz width is achieved. We have performed a same closed-loop operation on ground and obtain a frequency stability of $3.2 \times 10^{-12}/\tau^{1/2}$ by comparing to an H maser. According to the results, the short-term frequency stability in orbit is estimated to be better than $3.0 \times 10^{-13}/\sqrt{\tau}$ (*25*).

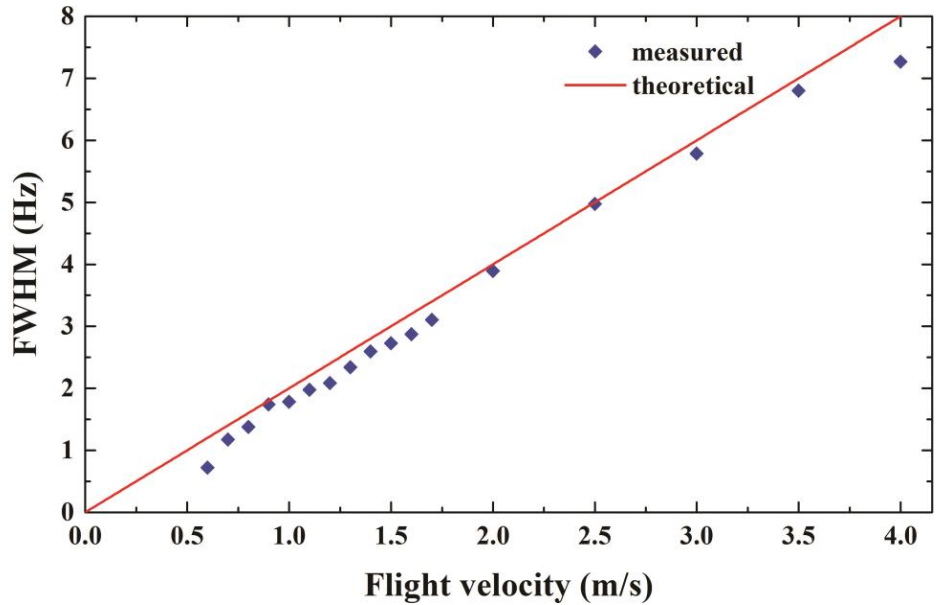

Fig. 4. The linewidth of central Ramsey fringe vs launching velocity of cold atoms in orbit. Due to the uniform motion of the cold atoms in microgravity, the linewidth is linearly related to the flight velocity.

Since launched, the CACES has been working in orbit for almost one year, and its performance is still kept as designed. The results of the CACES in orbit is useful for constructing next generation atomic time system in space as well as the quantum sensors based on cold atoms such as optical clocks, acceleration and rotation sensor. Furthermore, the experience of this mission, especially the data related to the orbital environments, is also of benefit to the extremely precise physics experiment such as the preparation of the ultra-cold quantum gases.

ACKNOWLEGEMENTS

We thank our colleagues at the Technology and Engineering Center for Space Utilization, Chinese Academy of Sciences, especially Profs. Yidong Gu, Guangheng Zhao, Min Gao, Congming Lü, Hongen Zhong and many others for their many discussions and supports, and thank Prof. Wenrui Hu of Institute of Mechanics, Chinese Academy of Science for his long-term supports on the CACES. We thank the China Manned Space Program for their supports. We also thank Dr. Yuanci Gao of the University of Electronic Science and Technology of China for his work on the microwave cavity, and the Nanopa Vacuum Co. for their help on the vacuum tube.